\documentclass[oldversion]{aa}  
\usepackage{graphicx}
\usepackage{natbib}
\usepackage{footmisc}

\usepackage{txfonts}
\newcommand{\lya}{\ifmmode {\rm Ly}\alpha \else Ly$\alpha$\fi}
\def\ffs{\hbox{$\,.\!\!^{\rm s}$}}
\def\ffas{\hbox{$\,.\!\!^{\prime\prime}$}}

\def\micron{$\mu$m}

\def\ergs{erg s$^{-1}$}

\def\msun{\ifmmode M_{\odot} \else M$_{\odot}$\fi}
\def\msunyr{\ifmmode M_{\odot} {\rm yr}^{-1} \else M$_{\odot}$ yr$^{-1}$\fi}
\def\zsun{\ifmmode Z_{\odot} \else Z$_{\odot}$\fi}
\def\lsun{\ifmmode L_{\odot} \else L$_{\odot}$\fi}

\def\mup{\ifmmode M_{\rm up} \else M$_{\rm up}$\fi}
\def\mlow{\ifmmode M_{\rm low} \else M$_{\rm low}$\fi}
\begin{document}
   \title{Millimeter observations of HCM\,6A, a gravitationally lensed \lya\ emitting galaxy at $z=6.56$
\thanks{Based on observations carried out with the IRAM 30 meter Telescope. IRAM is supported by INSU/CNRS (France), MPG (Germany) and IGN (Spain)}
}
   
\author{F.\,Boone\inst{1,2}
	\and
	 D.\,Schaerer\inst{3,4}
	\and
	R.\,Pell\'o\inst{4}
	\and
        F.\,Combes\inst{1}
	\and
	E.\,Egami\inst{5}
          }

   \offprints{frederic.boone@obspm.fr}

   \institute{
	LERMA, Observatoire de Paris, 
	61 Avenue de l'Observatoire, 
	F--75014 Paris, France
	\and
	Max-Planck-Institut f\"ur Radioastronomie, 
	auf dem H\"ugel 69, 
	D--53121 Bonn, Germany
	\and
	Geneva Observatory, Universit\'e de Gen\`eve,
	51 Chemin des Maillettes, 
	CH--1290 Sauverny, Switzerland
	\and
	Observatoire Midi-Pyr\'en\'ees, Laboratoire d'Astrophysique, UMR 5572, 
	14 Avenue E. Belin, 
	F--31400 Toulouse, France
	\and
	Steward Observatory, University of Arizona, 
	933 North Cherry Avenue, 
	Tucson, AZ 85721, USA
     }

   \date{Received date, 2007; accepted date, 2007}

 
  \abstract {The gravitationally lensed \lya\ emitting galaxy, { HCM\,6A}, detected
  by Hu et al.\ (2002) at $z$$=$$6.56$ behind the Abell\,370 cluster was
  observed with the MAMBO-2 array of bolometers at 1.2\,mm
  wavelength. The galaxy was not detected down to 1.08 mJy ($3
  \sigma$), but the depth of the observations and the lens
  amplification allow us to improve by approximately one order of
  magnitude previously published upper limits on far infrared emission
  of \lya\ emitting galaxies at this redshift.

  The following upper limits are derived from our observations
  assuming typical dust parameters: dust mass $<5.3\times 10^7$ \msun,
  IR luminosity $<2.1\times 10^{11}$ \lsun, and star formation rate
  $<35$ \msunyr. The observed restframe UV--optical--IR spectral
  energy distribution (SED) of this galaxy is compatible with that of
  normal spiral galaxies or blue compact dwarf galaxies. 
  { SEDs of prototypical ULIRGs, such as Arp 220,  are clearly excluded.}
Finally, we obtain an upper limit of $\la 2.1
  \times 10^{-2}$ \msunyr\ Mpc$^{-3}$ for the dust-obscured SFR
  density of \lya\ selected galaxies at $z \sim$ 6.6.}


   \maketitle
%

\section{Introduction}
\label{intro}
Nine of the 22 QSOs known to date at $z$$\sim$6
\citep{2006NewAR..50..665F, 2007arXiv0706.0914W} were detected at
millimeter wavelengths \citep{2003MNRAS.344L..74P,
2003A&A...406L..55B, 2004MNRAS.351L..29R, 2007arXiv0704.2053W}. The
infrared luminosities implied by the millimeter fluxes are typically
$10^{13}$\,L$_{\odot}$ and the dust masses are $>10^8$ \msun. In
addition, CO lines were detected in J1148+5251
\citep{2003A&A...409L..47B,2003Natur.424..406W} revealing a large
amount of dense molecular gas ($2.2 \times 10^{10}$ \msun) most
probably heated by a starburst rather than by an AGN. The Gunn-Peterson
trough seen in the optical spectrum of these objects indicate that
they are situated at the end of the epoch of reionisation
\citep{2001ARA&A..39...19L}. These observations imply that the process
of enrichment of the ISM is relatively advanced in these objects and
could have started at $z \ga 8$. This is in agreement with the results
of WMAP which place the first onset of star formation at $z \ga $11
\citep{2006astro.ph..3450P}.  These observations also put into
question the processes responsible for the dust formation. Indeed, at
the redshift of J1148+5251 the Universe was only $\sim 0.8$ Gyr old
and dust formation in quiescent winds of low mass stars would not have
been efficient enough to produce the observed masses.  Possible
sources of enrichment are SNII or pair instability SN
\citep[e.g.][]{2001MNRAS.325..726T,2004MNRAS.351.1379S}, which can
explain several observations in the QSOs at $z \sim 6$
\citep{2004MNRAS.349L..43V}. Quasar winds may also be a source of dust
production. To better understand the dust formation mechanisms and to
obtain a more global view of star formation in the early Universe it
is essential to observe $z>6$ non-quasar galaxies in the submillimeter
range. To date submillimeter observations of only two galaxies at
$z>6$ have been reported in the literature
\citep{2007ApJ...659...76W}, and each new observation brings important
constraints to the models.

The $z=6.56$ galaxy HCM6A was discovered by \citet{2002ApJ...568L..75H} 
in a narrow band search for gravitationally lensed \lya\ emitters behind the 
galaxy cluster Abell 370 at { $\alpha_{\rm J2000}=2^{\rm h} 39^{\rm m} 54\ffs 73$, $\delta_{\rm J2000}= -1^{\circ}33' 32\ffas 3$}. Several follow-up observations and analysis of this object
have been published. In particular, from an SED analysis
including constraints on its \lya\ emission,  \citet{2005MNRAS.362.1054S} suggested
that this galaxy could show a non-negligible extinction, with $A_v\sim 1$.
This result was later reinforced by the analysis of \citet{2005ApJ...635L...5C}
including also Spitzer observations, and more recently confirmed by \citet{2007MNRAS.376.1861F}.
If true, the {dust corrected} star formation rate of HCM 6A is SFR $\sim$ 11-41 \msunyr,
i.e.\ 2 to 8 times higher than previously thought,
and its luminosity is $L_{\rm bol} \sim (1-4)\times 10^{11}$ \lsun\
\citep{2005MNRAS.362.1054S} in the range of luminous infrared galaxies (LIRGs).
{This high SFR implies the presence of large amounts of gas, and therefore dust.}
This makes this strongly lensed galaxy \citep[magnification $\mu$=4.5 according to the model of ][]{1993A&A...273..367K} an interesting target to probe dust in very distant galaxies.

Here we report new submillimetric observations with the MAMBO-2
bolometer at the IRAM 30m telescope. These observations were
conducted in the frame of a larger program dedicated to near-IR and
submillimeter observations of $z>6$ galaxies behind lensing clusters,
as introduced by \citet{2006Msngr.125...20S}.

The paper is structured as follows: existing and new observations
of HCM 6A are described in Sect.\ 2. In Sect.\ 3 we estimate
the various physical parameters, and present the overall SEDs.
Our main conclusions are summarised in Sect.\ 4.
We assume a $\Lambda$-cosmology with H$_0$=70\,km\,s$^{-1}$\,Mpc$^{-1}$,
$\Omega_{\rm M}$=0.3 and $\Omega_{\Lambda}$=0.7.


\section{Observations}
\label{sec:1}

In Table \ref{t_obs} we have compiled the available observations
for HCM 6A from the literature. All fluxes have been converted to micro Jansky;
$3 \sigma$ limits are given for non-detections.

\begin{table}[htb]
\caption{Observations of HCM6A. $3 \sigma$ limits are given for non-detections.
Note that a nearby source, CBK4, has been detected by \protect\citet{2002AJ....123.2197C}
at 850 \micron. It is discussed in Sects.\ \ref{s_early} and \ref{s_dust}.
}
\begin{tabular}{lrllll}
\hline
\hline
Filter/     & wavelength  & flux & source \\
instrument & [$\mu$m]    & [$\mu$Jy] \\
\hline
V  & 0.55 & $<$ 0.06                & Hu et al.\ (2002) \\
R  & 0.65 & 0.02$\pm$0.01 \\
I  & 0.86 & $<$ 0.06   \\
Z  & 0.90 & 0.15$\pm$0.11 \\
J  & 1.25 & 0.25$\pm$0.12 \\
H  & 1.63 & 0.40$\pm$0.15 \\
K  & 2.12 & 0.16$\pm$0.13  \\
IRAC & 3.6     &  0.5$\pm$0.2       & Chary et al.\ (2006) \\
   & 4.5       & 1.25$\pm$0.3 \\
   & 5.8       & $<$ 2.7  \\
   & 8.0       & $<$ 3.  \\
JCMT   & 450   & $<$ 30000.         & Smail et al.\ (2002) \\
   & 850       & $<$ 1700    & Cowie et al.\ (2002) \\
MAMBO-2 & 1200 & $<$ 1080             & this paper \\
\hline
\label{t_obs}
\end{tabular}
\end{table}

\subsection{Earlier observations}
\label{s_early}
The optical and near-IR data (up to the $K^\prime$ band) is taken from the 
discovery paper of \citet{2002ApJ...568L..75H}.

At longer wavelength the IRAC Spitzer fluxes from Chary et al.\ (2006)
have been used. However, due to the presence of the nearby source in the south-west already seen in images of \citet{2002ApJ...568L..75H}, the IRAC photometry 
may be somewhat contaminated. For this reason and given the lower
spatial resolution, the MIPS 24 \micron\ data cannot be used.

The only references mentioning observations of Abell\,370 in the
sub/millimeter range are \citet{2002AJ....123.2197C} and
\citet{2002MNRAS.331..495S}. Both refer to SCUBA (JCMT) observations
and the first reference only mentions the detection of a source
(source No 4 in their Table\,3, hereafter CBK4) with a flux of
2.17$\pm0.57$\,mJy at 850\,$\mu$m and located $\sim 10$\arcsec\ to the West
of HCM\,6A. This emission likely originates from a foreground galaxy of the
cluster $\sim 5$\arcsec\ to the South-West of HCM\,6A. In addition this offset is larger than typical SCUBA astrometric uncertainties ($\sim$4$''$).
However, the astrometry becomes less accurate for weaker sources and we note that for another weak source in the field the positions given by \citet{2002AJ....123.2197C} and \citet{2002MNRAS.331..495S} could differ by 10\arcsec.
Indeed, the source \#2 of \citet{2002AJ....123.2197C}
detected at 5\,$\sigma$ might also be detected by
\citet{2002MNRAS.331..495S} but offset by $\sim$10$''$ (the source
J02400-0134). Hence,  it cannot be {definitely} excluded that the emission of CBK4 comes from HCM\,6A. The galaxy would be at the border of the field
covered by \citet{2002MNRAS.331..495S} but its flux would be below
their detection threshold.
In short, except otherwise stated, we retain the upper limits from SCUBA 
in Table \ref{t_obs} for HCM 6A (for the 850\,$\mu$m upper limit we take 3 times the uncertainty quoted for the flux of CBK4, i.e. 1.7\,mJy).

\subsection{New MAMBO-2 observations at 1.2 mm}
The observations were carried out at the IRAM 30m Telescope with the
MAMBO-2 array of 117 bolometers \citep{2002AIPC..616..262K}
during the
pool session in 2004/2005. It has a half-power spectral bandwidth of
80 GHz centered on $\sim$250\,GHz (1.2 mm). The beam size is
11$''$, thus, any emission originating from the foreground galaxy of the cluster  $\sim 5$\arcsec\ to the South-West of HCM\,6A would be dimmed by a factor of $\sim$2. The excellent weather conditions allowed us to reach an rms of
0.36\,mJy in 4.2 hours of integration in ON/OFF mode. The galaxy was
not detected, the flux measured is $-0.03\pm 0.36$\,mJy.
Our non-detection corresponds to a $3 \sigma$ upper limit of 1.08 mJy.

\section{Results}

The main derived quantities for HCM 6A are summarised in Table \ref{t_res}.
We now discuss them one by one.

\begin{table}[htb]
\caption{Derived quantities for HCM 6A assuming different dust temperatures 
$T_{\rm d}$, a magnification $\mu$$=$4.5, {a dust mass absorption coefficient $\kappa_{125}$$=$1.875\,m$^2$\,kg$^{-1}$, a dust emissivity index
$\beta$$=$1.5, and a high frequency dust spectral index $\alpha$$=$$2.9$}. All limits are $3 \sigma$ upper limits based on the 1.2\,mm flux upper limit for $T_{\rm d}$=18 and 36\,K and the 850\,$\mu$m  flux upper limit for  $T_{\rm d}$=54\,K.
For comparison, quantities estimated by \protect\citet{2005MNRAS.362.1054S}
are given at the bottom of the table. Note, however, that the latter
SFR estimates need to be multiplied by 2.55 to account for different
lower mass limits of the Salpeter IMF ($M_{\rm low}=1$ \msun\ instead of 
0.1 assumed for the SFR(IR) determination).}
\begin{tabular}{lrllll}
\hline
\hline
             & $T_{\rm d}=18$             & $T_{\rm d}=36$           & $T_{\rm d}=54$ \\
\hline
Dust mass $[$M$_{\odot}$$]$   & $<7.0\times 10^8$    & $<5.3\times 10^7$  & $<1.5\times 10^7$          \\
$L_{\rm FIR}$ $[$L$_{\odot}$$]$ & $<6.4\times 10^{10}$  & $<2.1\times 10^{11}$& $<5.4\times 10^{11}$  \\
SFR(IR) $[$M$_{\odot}$\,yr$^{-1}$$]$     & $<11$                & $<35$              & $<87$ \\
\hline
\multicolumn{2}{l}{$L_{\rm bol}$ $[$L$_{\odot}$$]$}  &  \multicolumn{2}{c}{$(1-4)\times 10^{11}$}  \\
\multicolumn{2}{l}{SFR$_{\rm UV}$($A_v=1$)   $[$M$_{\odot}$\,yr$^{-1}$$]$} & \multicolumn{2}{c}{11--41} \\
\multicolumn{2}{l}{SFR$_{\lya}$($A_v=1$)   $[$M$_{\odot}$\,yr$^{-1}$$]$} & \multicolumn{2}{c}{7--12} \\
\multicolumn{2}{l}{SFR$_{\rm UV}$($A_v=0$)  $[$M$_{\odot}$\,yr$^{-1}$$]$}   & \multicolumn{2}{c}{5.6} \\
\multicolumn{2}{l}{SFR$_{\lya}$($A_v=0$)  $[$M$_{\odot}$\,yr$^{-1}$$]$} & \multicolumn{2}{c}{0.4--0.8} \\
\hline
\label{t_res}
\end{tabular}
\end{table}

\subsection{Dust mass}
\label{s_dust}

\begin{figure}

  \includegraphics[width=0.95\columnwidth]{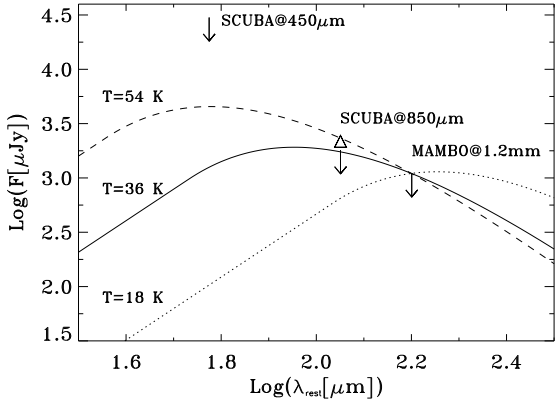}
\caption{Far infrared SEDs for dust temperatures of 54\,K (dashed), 36\,K (solid) and 18\,K (dotted) with $\beta$=1.5, $\alpha$=2.9 and such that $F_{\rm 1.2mm}$=1.1\,mJy. The triangle shows the possible detection at 850\,$\mu$m by \citet{2002AJ....123.2197C}, the arrows show the 3$\sigma$ upper limits as summarized in Table\,\ref{t_obs}.}
\label{fig:dust}       
\end{figure}
The upper limit on the MAMBO-2 flux allows us to compute an upper limit for the mass of dust using the fomula:
\begin{equation}
M_{\rm d}=\frac{S_{\nu}/\mu\,D^2_{\rm L}}{(1+z)\,\kappa_{\nu_{\rm r}}\,{B}_{\nu_{\rm r}}(T_{\rm d})}
\end{equation}
where $S_{\nu}$ is the flux upper limit, $\mu$ is the gravitational magnification factor, $D_{\rm L}$ is the luminosity distance,  $\kappa_{\nu_{\rm r}}$ is the mass absorption coefficient of the dust at $\nu_{\rm r}$, and ${B}_{\nu_{\rm r}}(T_{\rm d})$ is the intensity of a blackbody at $\nu_{\rm r}$ assuming isothermal emission from dust grains at temperature $T_{\rm d}$.  To compare with the upper limits reported by \citet{2007ApJ...659...76W} we also assume  $\kappa_{\rm 125\mu m}$=1.875\,m$^2$\,kg$^{-1}$ and  $\kappa_{\nu}\propto \nu^{\beta}$ with the emissivity index $\beta$=$1.5$ and we take $T_{\rm d}$=36\,K. This leads to a 3$\sigma$ upper limit of $M_{\rm d}$=$5.3\times 10^7$\,M$_{\odot}$.  This upper limit is one order of magnitude lower than the two upper limits given by  \citet{2007ApJ...659...76W} for two $z$$\sim$6.5 \lya\ emitters. This significant improvement in the dust mass upper limit of a \lya\ emitter at $z>$6 results from the high sensitivity of the observations and the strong gravitational amplification of the source.

The dust mass estimate is highly sensitive to the temperature assumed, the upper limit becomes  $2.0\times 10^7$\,M$_{\odot}$ for $T_{\rm d}$=54\,K and  $7.0\times 10^8$\,M$_{\odot}$ for $T_{\rm d}$=18\,K.  In addition, the dust  at high redshift is thought to result mainly from condensation of grains in SN ejecta and its emissivity might differ from that of the local Universe. With the emissivity introduced by \citet{2007MNRAS.378..973B} for dust formed in SN ejecta we obtain 65\% lower values for the upper limits on the dust mass. {Uncertainties in the emissivity index must also be considered. According to \citet{2007ApJ...662..284Y} the median value for LIRGs is $\beta$=1.6$\pm$0.3. Taking the higher value, i.e. $\beta$=1.9, increases the dust mass upper limits by 10\%.}
 However, despite the uncertainties, this upper limit {definitely} excludes the  dust masses found in the high-$z$ quasars and submillimeter galaxies (SMGs). It is closer to the mass of dust observed in local Universe galaxies \citep{2005MNRAS.364.1253V}.

If the emission seen in the 850\,$\mu$m map by \citet{2002AJ....123.2197C} is real and if the flux effectively originates from HCM6A (cf.\ above discussion about CBK4), it would imply a dust mass  $M_{\rm d}$=$6.5\times 10^7$\,M$_{\odot}$, assuming $T_{\rm d}=36$ K. Taking into account the uncertainties this is compatible with our upper limit at the same temperature. Our upper limit would however imply that the dust cannot be colder than 36\,K (see Fig.\,\ref{fig:dust}). If, as is most likely, the emission seen in their map does not originate from HCM6A, our 1.2\,mm dust mass upper limit improves their 850\,$\mu$m upper limit for dust colder than 36\,K. For a higher dust temperature their upper limit is more stringent, e.g. at 54\,K it would imply $M_{\rm d}$$<$$1.5\times 10^7$\,M$_{\odot}$. 

In the following and as summarized in Table\,\ref{t_res} we allow for an uncertainty of 50\% on the dust temperature by considering the three values, 18, 36 and 54\,K. When refering to the 2 lowest temperatures we {implicitly} use our 1.2\,mm flux upper limit and for the highest temperature we use the 850\,$\mu$m upper limit from \citet{2002AJ....123.2197C}. {The errors  seem to be dominated by the uncertainty on the temperature rather than the emissivity index. In addition, if, as proposed by  \citet{2007ApJ...662..284Y}, the emissivity index is negatively correlated to the temperature, then the variations of both parameters would compensate. To simplify the error analysis we therefore consider a range in temperature only and we assume that the resulting errors are good approximations to the errors due to all dust parameter uncertainties.}

\subsection{Far infrared luminosity}

As \citet{2007ApJ...659...76W} we compute the total far infrared  (FIR) luminosity assuming a modified black-body spectral energy distribution (SED) with a dust temperature of 36\,K and a dust emissivity of $\beta$=1.5 and replacing the Wien's exponential decrease shortward of 53\,$\mu$m by a power {law} with spectral index $\alpha$=2.9 to better agree with observed SEDs \citep{2003MNRAS.338..733B}. Our 1.2\,mm 3$\sigma$ flux upper limit corresponds to $L_{\rm FIR}$=1.9$\times$10$^{11}$\,\lsun, which is one order of magnitude lower than the FIR luminosities usually measured in the SMGs comparable to the local Universe ultraluminous infrared galaxies \citep{2004ApJ...614..671C, 2005ApJ...622..772C}.

This leads to an upper limit on the  $L_{\rm FIR}/ L_{\rm UV}$ ratio of $\sim$2. This is clearly within the range of values exhibited by Lyman-break galaxies of  $L_{\rm FIR}/ L_{\rm UV}<$10 \citep{2000ApJ...544..218A}. 
Using the fits of \citet{2005MNRAS.360.1413B} our $L_{\rm FIR}/ L_{\rm UV}$ limit translates
{into a dust attenuation in the far-ultraviolet (i.e. $\lambda$$=$1350--1750\,$\AA$)} $A_{\rm FUV} \la 0.9$, compatible with, but higher than the extinction estimated 
earlier from SED fits in the literature \citep[][]{2005MNRAS.362.1054S, 2005ApJ...635L...5C,2007MNRAS.376.1861F}.

The Ly$\alpha$ luminosity, $L_{\rm Ly\alpha} =1.9 \times 10^{42}$ \ergs, is $>0.3$\% of the bolometric luminosity, which seems to follow the trend noted for \lya\ blobs and SCUBA
galaxies by \citet{2005MNRAS.363.1398G} and \citet{2007ApJ...659...76W}.

\subsection{Star formation rate}
\label{s_sfr}

The upper limit on the star formation rate (SFR) derived from this FIR
luminosity with the relation from \citet{2003ApJ...586..794B} assuming
a Salpeter IMF from 0.1 to 100 \msun\ is SFR $< 35$ M$_{\odot}$\,yr$^{-1}$. 
This upper limit is a factor 8--20 lower than that of \citet{2007ApJ...659...76W} 
and \citet{2007ApJS..172..518C} for \lya\ emitters at 
$z \sim$ 5.7--6.6.

Although the SFR depends strongly on the dust temperature (the SFR
upper limit becomes 87\,M$_{\odot}$\,yr$^{-1}$ for $T_{\rm d}$=54\,K
and 11\,M$_{\odot}$\,yr$^{-1}$ for $T_{\rm d}$=18\,K as indicated in
Table \ref{t_res}) this upper limit definitely excludes star formation
rates observed in SMGs in excess of $\sim$1000
\,M$_{\odot}$\,yr$^{-1}$.  Our upper limit also excludes the high SFR
suggested for HCM 6A by \citet{2005ApJ...635L...5C}, who attributed an
apparent flux excess at 3.6 \micron\ to H$\alpha$ emission.
The SFR limit is slightly lower but still compatible within the
uncertainties with the {estimates of the dust corrected SFR from 
\citet{2005MNRAS.362.1054S} listed 
also in Table \ref{t_res}, and based on the SED fits to UV--optical fluxes.}
Note that for comparison the latter SFR values should be multiplied 
by a factor 2.55 to account for the different lower mass limit of 
the IMF ($M_{\rm low}=1$ \msun).

\begin{figure}[htb]
  \includegraphics[width=0.95\columnwidth]{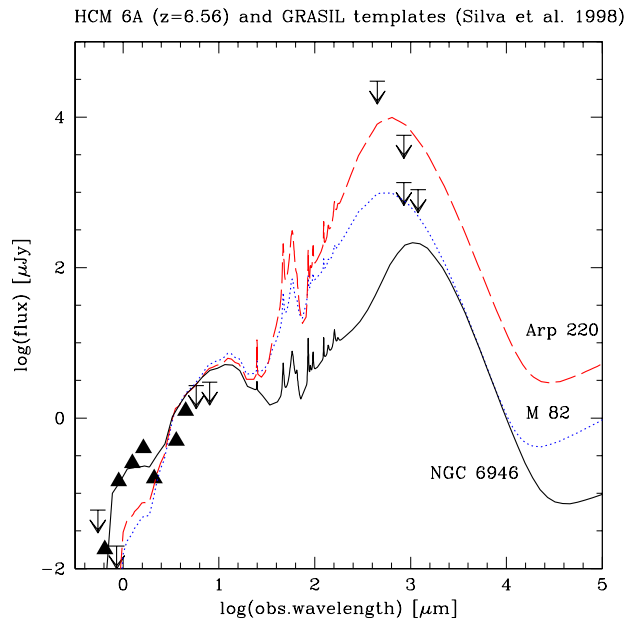}
\caption{Comparison of the observed SED of HCM 6A {(the data points are listed in Table\,\ref{t_obs})} with GRASIL templates 
for Arp 220, M82, and NGC 6946 from \citet{1998ApJ...509..103S} (from top to
bottom in the IR, normalised arbitrarily at 4 \micron). 
The observations in the near-IR and/or sub/millimeter range
rule out Arp 220 and M82 like templates, but are compatible with normal
galaxies as NGC 6946.}
\label{fig:1}       
\end{figure}

\subsection{Star formation rate density}

The gravitational magnification allows us to probe a galaxy with a
lower Ly$\alpha$ luminosity, i.e.  $L_{\rm Ly\alpha}$=$2\times
10^{42}$\,erg\,s$^{-1}$ instead of $7.5\times 10^{42}$\,erg\,s$^{-1}$
for the average of the 2 galaxies discussed by \citet{2007ApJ...659...76W}.
According to the luminosity function (LF) of
\citet{2006ApJ...648....7K} the cumulative number density is at least
10 times higher at this luminosity and our upper limit on the SFR
should therefore be more representative for the whole population of
\lya\ emitters.

To determine an upper limit for the infrared-measured SFR density (SFRD)
we assume that the number density is given by the cumulative 
number density at $L_{\rm Ly\alpha}$=$2\times 10^{42}$\,erg\,s$^{-1}$, which
is $N(>L) \sim 7.\times 10^{-4}$ Mpc$^{-3}$. 
For $T_{\rm d}=36$ K, we then obtain SFRD(IR)$ < 2.4 \times 10^{-2}$
\msunyr\ Mpc$^{-3}$, 
with a possible uncertainty of a factor of $\sim 3$
due to the unknown dust temperature (cf.\ Table \ref{t_res}). 
Integration over the LF of \citet{2006ApJ...648....7K} down to $0.5
L_\star$ (assuming $\alpha=-1.5$, $\log L_\star=42.6$, and that
$L_{\rm Ly\alpha}$ has the same proportionality with SFR at all luminosities) 
would imply an SFRD higher by a factor 2.2.

Incidentally our upper limit for SFRD(IR) is very similar to the value
estimated by \citet{2007ApJ...659...76W}. 
However, we note that these
authors adopted a lower value for the cumulative number density drawn from the
spectroscopically confirmed sample of \citet{2006ApJ...648....7K},
whereas we use the LF from the complete $z=6.5$ sample.
The SFRD(IR) limit can be regarded as a first estimate of the
dust-obscured SFR density of \lya\ selected galaxies at $z \sim$ 6.6.
More observations are clearly needed to improve this estimate.

For comparison, integration of the complete \lya\ LF, with the same
assumptions as above, yields SFRD(\lya)$=$$2.7 \times 10^{-3}$ \msunyr\
Mpc$^{-3}$ at $z=6.6$ assuming a standard \lya\ SFR conversion of 
SFR(\lya)$=9.1 \times 10^{-43} L_{\rm Ly\alpha}$ \ergs. 
For well known reasons (dust and radiation transfer effects, partial IGM
tranmission) this should represent a lower limit of the SFRD of \lya\
emitters at this redshift.

\begin{figure}[htb]
  \includegraphics[width=0.95\columnwidth]{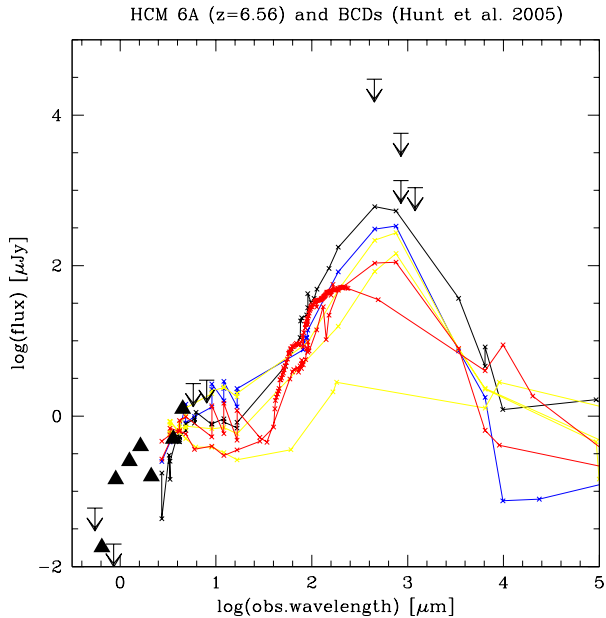}
\caption{Comparison of the observed SED of HCM 6A with templates 
of low metallicity Blue Compact Dwarf galaxies from \citet{2005A&A...434..849H}
normalised arbitrarily at 4 \micron. The SED of BCDs, including the 
rest-frame UV part not shown here, are compatible with HCM 6A.}
\label{fig:2}
\end{figure}

\subsection{Comparison to empirical SEDs}
In Fig.\ 2 we compare the observed SED (cf.\ Table \ref{t_obs}) with
semi-empirical templates from the GRASIL models of \citet{1998ApJ...509..103S}.
Shown are the templated of Arp220, a prototypical ULIRG, the prototypical starburst galaxy M82, 
and NGC 6946, an Scd galaxy.
Figure 3 shows a comparison with the observed SEDs of low metallicity blue compact dwarf galaxies
from Hunt et al.\ (2005).
{Clearly the observed millimeter to IR flux ratio excludes the ULIRG SED, and possibly
also the dusty starburst (M82). This is in agreement with the moderate extinction 
($A_V \la 1$) estimated from the fits to the (restframe) 
UV--optical observations.}
On the other hand, the SED of HCM 6A is compatible with that of normal galaxies, like NGC 6946, 
and with the SEDs of blue compact dwarf galaxies (BCD).
The present data does not allow us to distinguish between these two different types of SEDs, 
which mostly differ in their average dust temperature (dust being
hotter in BCDs). However, HCM 6A resembles more likely BCDs with higher surface brightness 
(or SFR surface density) than spiral galaxies. This is also found more generally for 
high-$z$ \lya\ emitters \citep{2007ApJ...660...47D}. 
In fact, local BCDs have SFR $\sim$0.2--3 \msunyr, quite similar to the lower limit determined for HCM 6A.

From the empirical SEDs shown in Figs.\ 1 and 2 we see that the millimeter flux
may be close to a detection. In any case this object will be an interesting target for ALMA,
which should be able to characterise the dust properties of this faint distant galaxy.
\section{Discussion and conclusions}

Quantifying the dust content in distant galaxies is important to
constrain dust formation mechanisms in the early Universe and to 
determine cosmological properties such as the star formation rate 
density (SFRD) at high redshift.
Strong gravitational lensing provides a unique access to the faintest
observable sources probing the galaxy luminosity function down to more 
representative objects.
Here we have used this effect to measure the millimeter emission of a 
faint lensed \lya\ emitter, suspected to suffer from a non-negligible 
amount of extinction in earlier studies.

Our MAMBO-2 observations of the $z=6.56$ galaxy HCM 6A yield
an upper limit of  1.08 mJy ($ 3 \sigma$) at 1.2 mm.
For typical dust parameters, this translates into upper limits
of $\la 5 \times 10^7$ \msun\ for the dust mass, 
$\la 2 \times 10^{11}$ \lsun\ for the infrared luminosity, and
$\la 35$ \msunyr\ for the IR star formation rate (SFR).
Thanks to strong gravitational lensing and to excellent weather conditions 
these limits improve recent sub-mm measurements of two $z \sim 6.5$
\lya\ emitters by approximately one order of magnitude.

Assuming a dust production of $\sim$ 0.1 \msun\ per type II supernova \citep{2007MNRAS.378..973B} a maximum age 
of $\sim 1/2 t_{H} \approx 4 \times 10^8$ yr, would require a constant 
SFR $\sim 60$ \msunyr\ (for $M_{\rm low} = 1$ \msun) to produce 
$5 \times 10^7$ \msun\ of dust.
Hence our results are marginally consistent with current estimates for the 
dust production. Reducing the large age uncertainty of HCM 6A,
and lowering the upper limit for its dust mass will allow to place
more severe constraints on dust formation scenarios.

Our non-detection of dust does not exclude the presence of dust
in HCM 6A. In fact the SFR(IR) upper limit of $\la 35$ \msunyr,
with a typical uncertainty of a factor 3 {in either direction}, comes close to but 
remains compatible with the UV SFR of $\sim$ 28--105 \msunyr\
obtained by \citet{2005MNRAS.362.1054S}\footnote{The SFR(UV) value 
quoted here has been converted to the same ``standard'' IMF adopted 
for the SFR(IR); cf.\ Sect.\ \ref{s_sfr}} for an extinction of $A_V \sim 1$.
Its is tempting then to speculate that the IR emission from this 
object could soon be detected with somewhat deeper sub/millimeter 
observations. 
However, another explanation for the apparently red SED of the
Ly$\alpha$ emitter HCM6A may also be a composite stellar population
with little or no dust, as discussed by \citet{2005MNRAS.362.1054S}.  
Deeper sub/millimeter observations of lensed high-$z$ galaxies 
including HCM6A should therefore soon be able to match the constraints from
shorter wavelength observations and provide more detailed
constraints on dust in the early Universe.

Thanks to gravitational lensing the \lya\ emitter observed
here corresponds, with $L_{\rm Ly\alpha}=1.9 \times 10^{42}$ \ergs, 
to a $\sim 0.5 L_\star$ galaxy according to the $z \sim 6.5$
\lya\ luminosity function (LF) of \citet{2006ApJ...648....7K},
at the faintest limits of their blank field survey limits.
Using our millimeter non-detection and their LF we determine an upper 
limit for the infrared-measured SFRD of $ < 2.4 \times 10^{-2}$
\msunyr\ Mpc$^{-3}$ {with an uncertainty of a factor 3  in either direction}. This can be regarded as an estimate of the
dust-obscured SFR density of \lya\ selected galaxies at $z \sim$ 6.6.
More observations will be needed to improve this estimate.
Gravitational lensing and new upcoming facilities should 
allow significant progress in the near future.

\begin{acknowledgements}
We thank Raffaella Schneider and Roberto Maiolino for interesting comments
on an earlier version of the manuscript.
Support from {\em ISSI} (International Space Science Institute) in Bern for an 
``International Team'' is gratefully acknowledged. 
This work was supported by the Swiss National Science Foundation,
the French {\it Centre National de la Recherche Scientifique},
and the French {\it Programme National de
Cosmologie} (PNC) and {\it Programme National de Galaxies} (PNG).

\end{acknowledgements}

\bibliographystyle{aa}      
\bibliography{hcm6a}   

\end{document}